\PassOptionsToPackage{pdftex}{graphicx}
\documentclass[mathpazo]{cicp}

\usepackage{bbm}
\usepackage{url}
\newcommand{\ii}{\mathbbm{i}}

\renewcommand\({\left(}
\renewcommand\){\right)}
\renewcommand\[{\left[}
\renewcommand\]{\right]}
\newcommand{\CLASS}{\textsc{class}}

\newcommand{\mpmath}{\textsc{mpmath}}
\newcommand{\Mathematica}{\textsc{Mathematica}}

\newcommand{\diff}[2]{\frac{ \textrm{d} #1}{\textrm{d} #2}}
\newcommand{\difff}[2]{\frac{ \textrm{d}^2 #1}{\textrm{d} #2^2}}

\newcommand{\cotK}{\cot _K}
\newcommand{\sinK}{\sin _K}

\newcommand{\atan}{\,\text{atan}}
\newcommand{\atanh}{\,\text{atanh}}

\title{Computation of hyperspherical Bessel functions}

\author[T.~Tram]{Thomas Tram\corrauth}
\address{Department of Physics and Astronomy, Aarhus University, Ny Munkegade 120, 8000 Aarhus C, Denmark}
\email{{\tt thomas.tram@phys.au.dk} (T.~Tram)}

\begin{document}

\begin{abstract}
In this paper we present a fast and accurate numerical algorithm for the computation of hyperspherical Bessel functions of large order and real arguments. For the hyperspherical Bessel functions of closed type, no stable algorithm existed so far due to the lack of a backwards recurrence. We solved this problem by establishing a relation to Gegenbauer polynomials. All our algorithms are written in C and are publicly available at Github [\url{https://github.com/lesgourg/class_public}]. A Python wrapper is available upon request.
\end{abstract}


\maketitle

\section{Introduction}
Hyperspherical Bessel functions are generalisations of spherical Bessel functions. They are needed for the computation of the anisotropy spectrum of the Cosmic Microwave Background (CMB) radiation for models with spatial curvature. While the differential equation can of course be integrated using ODE-solvers, no available high accuracy implementation of hyperspherical Bessel functions existed before this work. This was partly due to a problem of using backwards recurrence for hyperspherical Bessel functions of positive curvature.

%
\section{Analytic properties of hyperspherical Bessel functions}
\subsection{Definition}
The hyperspherical Bessel functions $\Phi^\nu_l(\chi)$ are the radial part of the eigenfunctions of the Laplacian on a 3 dimensional manifold of constant curvature. The sign of the curvature is denoted by $K$ and $\Phi^\nu_l(\chi)$ can be written as
\begin{equation}
\Phi^\nu_l(\chi) = \frac{u^\nu_l(\chi)}{r(\chi)},
\label{eq:definition_phi}
\end{equation}
where $u^\nu_l(\chi)$ is the solution regular at $\chi=0$ of the linear second-order differential equation
\begin{equation}
\difff{u^\nu_l}{\chi} = \[ \frac{l(l+1)}{r(\chi)^2} - \nu^2\] u^\nu_l(\chi).
\label{eq:u_equation}
\end{equation}
The dependence on geometry is encoded in the function $r(\chi)$ given by
\begin{equation}
r(\chi) = \sinK(\chi) \equiv \left\{
\begin{array}{ll}
\sinh \chi 	&K=-1\\
\chi 				&K=0\\
\sin \chi 	&K=1\\
\end{array}
\right. 
\end{equation}
Note that in flat space ($K=0$), one can do the transformation $z = \nu \chi$ and multiply through by $r(\chi)^2=\chi^2$ to transform equation~\eqref{eq:u_equation} into the Ricatti-Bessel equation, in which case $\Phi^\nu_l(\chi)=j_l(\nu\chi)$. 
\subsection{Recursive solutions}
The solutions to equation~\eqref{eq:u_equation} are known~\cite{opac-b1090215} and they can be written recursively as
\begin{equation} \label{eq:recursivesolution}
y^\nu_l(\chi) = \left\{
\begin{array}{rl}
\sinh^{l+1} \chi \( \frac{1}{\sinh \chi} \diff{}{\chi} \) ^{l+1} \(C_1 \cos \nu \chi + C_2 \sin \nu \chi \) & K=-1\\
\chi^{l+1} \( \chi \diff{}{\chi} \)^{l+1} \(C_1 \cos \nu \chi + C_2 \sin \nu \chi \) &K=0\\
\sin^{l+1} \chi \( \frac{1}{\sin \chi} \diff{}{\chi} \) ^{l+1} \(C_1 \cos \nu \chi + C_2 \sin \nu \chi \) &K=1\\
\end{array} 
\right..
\end{equation}
The solution becomes regular at $x=0$ by putting $C_2=0$ which can easily be proven by induction. By letting $\chi\rightarrow -\chi$ in the solutions of equation~\eqref{eq:recursivesolution}, we find that $y^\nu_l(-\chi) = (-1)^l y^\nu_l(\chi)$ or equivalently 
\begin{equation}\label{eq:evenodd}
\Phi^\nu_l(-\chi) = (-1)^l \Phi^\nu_l(\chi).
\end{equation}
Thus, the hyperspherical Bessel functions are even or odd depending on $l$ which is well known for normal spherical Bessel functions.
\subsection{Solutions in terms of Legendre functions}
The hyperspherical Bessel functions for $K=\pm1$ can be expressed as Legendre functions by a change of variables~\cite{RevModPhys.39.862,Abbott:1986ct}. The solutions which are regular at $x=0$ then reads
\begin{equation}\label{eq:phiLegendre}
\Phi^\nu_l(\chi) = \left\{
\begin{array}{ll}
\sqrt{\frac{\pi N^\nu_l}{2\sinh \chi}} P^{-1/2-l}_{-1/2+\ii\nu}(\cosh\chi) 	&K=-1\\
j_l(\nu \chi) 				&K=0\\
\sqrt{\frac{\pi M^\nu_l}{2\sin \chi}} P^{-1/2-l}_{-1/2+\nu}(\cos \chi) 	&K=1\\
\end{array}
\right.,
\end{equation}
where we have restricted ourselves to $\chi\geq0$ and also $\chi\leq \pi$ for $K=1$. By using equation~\eqref{eq:evenodd}, we can extend the solutions of equation~\eqref{eq:phiLegendre} to the whole real axis. For $K=1$ we are also using the $2\pi$-periodicity. The normalisation constants
\begin{equation}
N^\nu_l \equiv \prod_{n=1}^l{(\nu^2+n^2)}, \qquad
M^\nu_l \equiv \prod_{n=1}^l{(\nu^2-n^2}),
\end{equation}
have been chosen such that the $K\neq0$ hyperspherical Bessel functions are normalised similarly to the spherical Bessel functions~\cite{Abbott:1986ct}. 

Note that our definition of the Legendre functions follows the previous papers~\cite{RevModPhys.39.862,Abbott:1986ct,Kosowsky:1998nc}, so it is slightly inconsistent. In the $K=1$ case, $P_\alpha^\beta(x)$ denotes Ferrer's function of the first kind. In Mathematica, this function is the default 'type 1' of Legendre function. When it is extended to the whole complex plane it has branch cuts at $(-\infty,-1)$ and $[1,\infty)$, and is denoted by 'type 2' in Mathematica. For $K=-1$, $P_\alpha^\beta(x)$ is the Legendre function called 'type 3' by Mathematica and has a single branch cut $(-\infty,1]$ when extended to the complex plane. In terms of the Gauss hypergeometric function we have
\begin{align}
P_\alpha^\beta(x) &= \( \frac{1+x}{1-x} \)^\frac{\beta}{2} \frac{1}{\Gamma(1-\beta)} {\, _2F_1}\(\alpha+1,-\alpha,1-\beta,\frac{1}{2}-\frac{1}{2}x\), & K&=1, \\
P_\alpha^\beta(x) &= \( \frac{x+1}{x-1} \)^\frac{\beta}{2} \frac{1}{\Gamma(1-\beta)} {\, _2F_1}\(\alpha+1,-\alpha,1-\beta,\frac{1}{2}-\frac{1}{2}x\), & K&=-1.
\end{align}
\subsection{Special properties for $K=1$}
The $K=1$ case is special, since $\Phi^\nu_l(\chi)$ must satisfy an additional boundary condition at $\chi=\pi$ where $\frac{l(l+1)}{\sin^2\chi}\rightarrow \infty$. The limit $\chi \rightarrow \pi$ corresponds to $x \equiv \cos\chi \rightarrow -1^+$ in the argument of the Legendre function. We use the following formula\footnote{\url{http://dlmf.nist.gov/14.9.E9}} to relate the limit $x\rightarrow -1^+$ to $x\rightarrow 1^-$:
\begin{equation}
P_\nu^\mu(-x) = \cos\[(\nu+\mu)\pi\] P_\nu^\mu (x) - \frac{2}{\pi} \sin\[(\nu+\mu)\pi \] Q_\nu^\mu(x).\label{eq:Legendre_connection}
\end{equation}
Since the limit $x\rightarrow 1^-$ is equivalent to $\chi\rightarrow 0$, we already know that $P_\nu^\mu (x)$ is regular while $Q_\nu^\mu(x)$ diverges, but this behaviour can also be checked from the limiting forms\footnote{\url{http://dlmf.nist.gov/14.8.E6}} of $P_\nu^\mu(x)$ and $Q_\nu^\mu(x)$. For the boundary condition to be regular at $\chi=\pi$ we must then have the sine function in equation~\eqref{eq:Legendre_connection} to be identically zero,
\begin{equation}
\sin\[(\nu-l-1)\pi\]=0,
\end{equation}
so $\nu$ must be an integer. This restriction in possible solutions is equivalent to standing waves in a cavity or the quantum mechanical quantisation of energy in a potential well. Using equation~\eqref{eq:Legendre_connection} with integer $\nu$, we find a corresponding connection formula for $\Phi_l^\nu(\chi)$:
\begin{align}
\Phi^\nu_l(\pi-\chi) &= \sqrt{\frac{\pi M^\nu_l}{2\sin (\pi-\chi)}} P^{-1/2-l}_{-1/2+\nu}\(-\cos(\chi)\), \label{eq:symmetry}\\
&= \sqrt{\frac{\pi M^\nu_l}{2\sin (\chi)}} \cos\[(\nu-l-1)\pi\] P^{-1/2-l}_{-1/2+\nu}\(\cos(\chi)\), \nonumber \\
&= (-1)^{\nu-l-1} \Phi^\nu_l(\chi), \qquad (K=1) \nonumber
\end{align}
which shows that $\Phi^\nu_l(\chi)$ is symmetric (anti-symmetric) around $\chi=\frac{\pi}{2}$ for $\nu-l-1$ even (odd)\footnote{This symmetry was noted by~\cite{Kosowsky:1998nc}, but his equation is incorrect: it holds for the Legendre function, not for $\Phi$ as indicated. The same erroneous formula is also found in~\cite{Abbott:1986ct,RevModPhys.39.862}.}. Equation~\eqref{eq:evenodd} combined with equation~\eqref{eq:symmetry} allows us to restrict ourselves to the region $\[0,\frac{\pi}{2}\]$ in the $K=1$ case.

Another subtlety is the allowed range of $l$ for a given $\nu$. Consider the right hand side of equation~\eqref{eq:u_equation}. If $l\geq \nu$, the coefficient $\[ \frac{l(l+1)}{r(\chi)^2} - \nu^2\]$ will always be positive, and there will be no oscillatory region. Thus, only the trivial solution $\Phi^\nu_l(\chi)=0$ is allowed in this case. (In quantum mechanics, is is well-known that all states must have energy $\nu^2$ greater than the minimum of the potential.)  
\subsection{Relation to the Gegenbauer polynomials}
The Legendre function is related to the Gegenbauer function $C_\alpha^{(\beta)}$ by the identity\footnote{\url{http://dlmf.nist.gov/14.3.E21}}
\begin{equation}
P_\alpha^\beta(x) = \frac{2^\beta\Gamma(1-2\beta)\Gamma(\alpha+\beta+1)}{\Gamma(\alpha-\beta+1) \Gamma(1-\beta)(1-x^2)^{\beta/2}}C_{\alpha+\beta}^{(\frac{1}{2}-\beta)}(x). \label{eq:legendregegen}
\end{equation}
Surprisingly, when $\nu$ is a positive integer, the order of the Gegenbauer function $\alpha+\beta$ also becomes a positive integer and the Gegenbauer function reduces to the Gegenbauer polynomial. We find explicitly
\begin{align} 
\Phi^\nu_l(\chi) &= \sqrt{\frac{\pi M^\nu_l}{2\sin \chi}} P^{-1/2-l}_{-1/2+\nu}(\cos \chi) \nonumber \\
&= \sqrt{\frac{\pi M^\nu_l}{2\sin \chi}} \frac{2^{-\frac{1}{2}-l} \Gamma(2+2l)\Gamma(\nu-l)}{\Gamma(\nu+l+1)\Gamma(\frac{3}{2}+l) (\sin^2\chi)^{-\frac{1}{2}(\frac{1}{2}+l)}}C_{\nu-l-1}^{(l+1)}(\cos \chi) \nonumber \\
&= 2^l l! \sqrt{\frac{(\nu-l-1)!}{\nu(\nu+l)!}} \sin^l(\chi) C_{\nu-l-1}^{(l+1)}(\cos \chi)  \label{eq:Gegenbauer}
\end{align}
Let us emphasise that this relationship between $K=1$ hyperspherical Bessel functions and Gegenbauer poynomials is a new result to the best of our knowledge. In deriving equation~\eqref{eq:Gegenbauer} we have used 
\begin{equation}
M_l^\nu = \( \prod_{n=1}^l{\( \nu-n \)} \) \( \prod_{n=1}^l{\( \nu+n \)} \)
= \frac{(\nu-1)!}{\nu-l-1)!} \frac{(l+\nu)!}{\nu!} =\frac{(\nu+l)!}{\nu(\nu-l-1)!}, \nonumber
\end{equation}
and the duplication formula for the gamma function $\Gamma(2(l+1))$.

Because the Gegenbauer polynomials are easy to compute in a stable fashion, equation~\eqref{eq:Gegenbauer} can be used to create a compact method for the computation of the $K=1$ hyperspherical Bessel functions. However, for our application we will need to compute the hyperspherical Bessel functions for fixed $\nu$ and many values of $l$. For that purpose the direct implementation is sub-optimal, since recurrence in the Gegenbauer polynomials is effectively a recurrence in $\nu$. This means that most points in the recurrence sequence will be thrown away. We will instead use equation~\eqref{eq:Gegenbauer} to compute a starting point for the backwards recurrence in $l$ as demonstrated later in section~\ref{sec:CF1K}.
\section{Computationally efficient algorithms}
\subsection{Existing software packages}
\begin{figure}[htp]
\begin{centering}
\includegraphics[width=\textwidth]{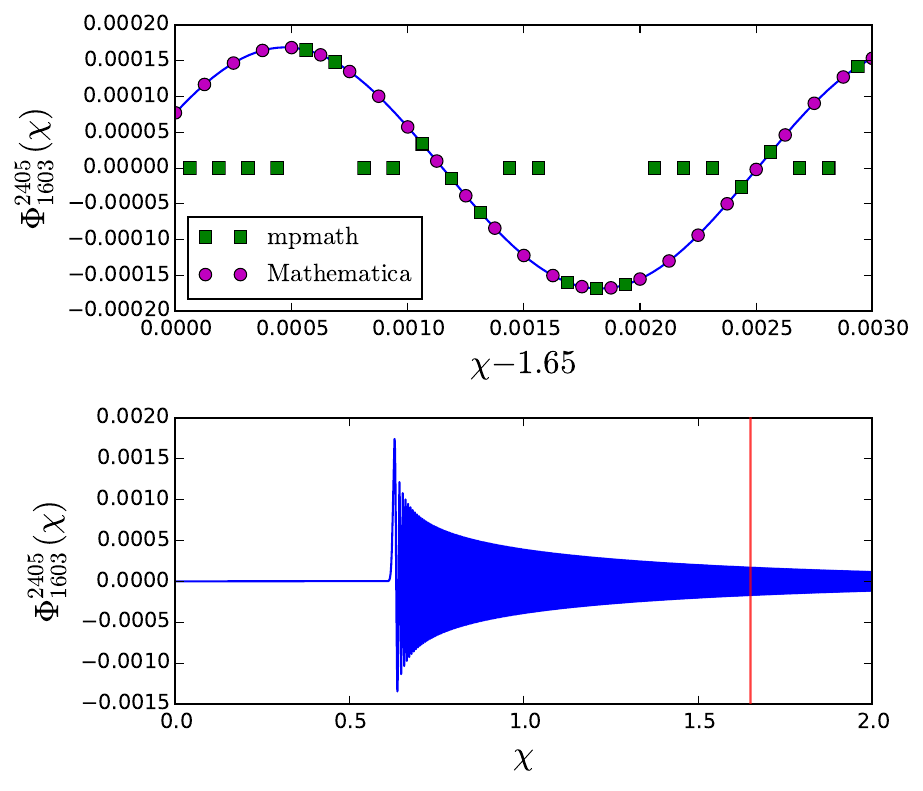}
\label{fig:hyperfail}
\caption{One example of a region in the $K=-1$ parameter space where existing software packages have trouble. The \mpmath{} implementation spends 22 seconds on computing 24 points. The 13 points not located on the curve have failed to converge. \Mathematica{} spends 4 CPU-hours on computing the 25 points shown. The $\chi$-region of the top plot is shown as a red stripe in the bottom plot.}
\end{centering}
\end{figure}
The open source python library \mpmath{}~\cite{mpmath} has a reasonable fast and stable implementation of the Gauss hypergeometric function for complex orders. There are two ways of implementing hyperspherical Bessel functions using \mpmath{}, either by using the Legendre function or the hypergeometric function. For some parameters, one function would converge and the other would not and vice versa. For some points in the parameter space, neither method would converge as shown in figure~\ref{fig:hyperfail}. \Mathematica{} is able to converge on the correct value in this parameter space, but computing the 25 points in~\ref{fig:hyperfail} took 4 CPU hours. As a comparison, the method described in the next section computes $\Phi$ on 250,000 linearly spaced points in the same parameter range in 2.97 CPU seconds. For this choice of parameters, our implementation is then more than 6 orders of magnitude faster than \Mathematica{}.

\subsection{Recurrence relations}
The solutions in equation~\eqref{eq:phiLegendre} are of little use from the numerical point of view, since we do not have a general method for computing the Legendre functions at large order. The hyperspherical Bessel functions can also be expressed in terms of the Gauss hypergeometric function, but this has again no computational advantage. However, the functions satisfy the recurrence relations~\cite{Abbott:1986ct}
\begin{align}
\Phi^\nu_l(\chi) &= \frac{1}{\sqrt{\nu^2-Kl^2}} \left\{ (2l-1) \cotK \chi \Phi^\nu_{l-1}(\chi) - \sqrt{\nu^2-K(l-1)^2} \Phi^\nu_{l-2}(\chi) \right\}, \label{eq:phirec}\\
{\Phi^\nu_l}^\prime(\chi) &= l \cotK \chi \Phi^\nu_l(\chi) - \sqrt{\nu^2-K(l+1)^2} \Phi^\nu_{l+1}(\chi) , \label{eq:dphirec}
\end{align}
where we have used the notation
\begin{equation}
\cotK \chi \equiv \left\{
\begin{array}{ll}
\coth \chi 	&K=-1\\
\frac{1}{\chi} 		&K=0\\
\cot \chi 	&K=1\\
\end{array}
\right. 
\end{equation}
These recurrence relations form the basis of our method. For small $l$, the hyperspherical Bessel functions are given by the simple analytic formulae,
\begin{align}
\Phi^\nu_0(\chi) &= \frac{\sin (\nu \chi) }{\nu \sinK \chi}, \\
\Phi^\nu_1(\chi) &= \Phi^\nu_0(\chi) \frac{\cotK \chi - \nu \cot (\nu \chi)}{\sqrt{\nu^2-K}}.
\end{align}
This allows a forwards recurrence for part of the parameter space. However, since $\Phi^\nu_l(\chi)$ represents the minimal solution, forwards recurrence will be unstable for regions outside the classical turning point. 
\subsection{Backwards recurrence}
The solution is to use the recurrence backwards, so let us address the problem of initial values for the backwards recurrence. By defining
\begin{equation}
\alpha_l \equiv \frac{(2l+1)\cotK\chi }{\sqrt{\nu^2-K(l+1)^2}}, \qquad
\beta_l \equiv -\frac{\sqrt{\nu^2-Kl^2}}{\sqrt{\nu^2-K(l+1)^2}}, \qquad
y_l \equiv \Phi^\nu_l(\chi), 
\end{equation}
the recurrence relation in equation~\eqref{eq:phirec} takes the form
\begin{align}
y_{l+1} = \alpha_l y_l + \beta_l y_{l-1}.
\end{align}
Dividing through by $y_l$ and rearranging terms yields
\begin{equation}
-\frac{y_l}{y_{l-1}} = \frac{\beta_l}{\alpha_l-\frac{y_{l+1}}{y_l}},
\end{equation}
which can be iterated to give the continued fraction
\begin{equation}
-\frac{y_l}{y_{l-1}} = \frac{\beta_l}{\alpha_l+}\frac{\beta_{l+1}}{\alpha_{l+1}+} \cdots \frac{\beta_{l+j}}{\alpha_{l+j}+}\cdots. \label{eq:CF0}
\end{equation}
The continued fraction converges according to Pincherle's theorem~\cite{Gautschi:1967:CAT,Press:1992:NRC:148286} since $y_l=\Phi^\nu_l(\chi)$ is the minimal solution. By dividing~\eqref{eq:dphirec} by $\Phi^\nu_l(\chi)$ and using equation~\eqref{eq:CF0} we finally find
\begin{equation}
\frac{{\Phi^\nu_l}^\prime(\chi)}{\Phi^\nu_l(\chi)} = l\cotK \chi + \sqrt{\nu^2-K(l+1)^2} \left\{ \frac{\beta_{l+1}}{\alpha_{l+1}+}\frac{\beta_{l+2}}{\alpha_{l+2}+} \cdots \frac{\beta_{l+j}}{\alpha_{l+j}+}\cdots\right\}. \quad \text{(CF1)} \label{eq:CF1}
\end{equation}
CF1, equation~\eqref{eq:CF1}, is finally evaluated using the modified Lentz method~\cite{Lentz:76,Thompson1986490,Press:1992:NRC:148286}. 
\subsection{CF1 for $K=1$}\label{sec:CF1K}
For $K=1$ the iteration may not always converge because the restriction $l<\nu$ puts an upper limit on the number of iterations. In practice what happens is that the argument of the square root $\sqrt{\nu^2-K(l+j+1)^2}$ in equation~\eqref{eq:CF0} becomes less than zero. However, by using the relation to Gegenbauer polynomials in equation~\eqref{eq:Gegenbauer}, we can evaluate $\frac{{\Phi^\nu_l}^\prime(\chi)}{\Phi^\nu_l(\chi)}$ directly. From equation~\eqref{eq:Gegenbauer} we get
\begin{align}
{\Phi^\nu_l}^\prime(\chi) &=2^l l! \sqrt{\frac{(\nu-l-1)!}{\nu(\nu+l)!}} 
\sin^l(\chi) \[ l \cot\chi C_{\nu-l-1}^{(l+1)}(\cos \chi)-\sin\chi {C_{\nu-l-1}^{(l+1)}}^\prime(\cos \chi) \], \nonumber \\
&= \[l \cot \chi - \sin\chi\frac{{C_{\nu-l-1}^{(l+1)}}^\prime(\cos \chi)}{C_{\nu-l-1}^{(l+1)}(\cos \chi)} \] \Phi^\nu_l(\chi). 
\label{eq:CF1fromGegenbauer}
\end{align}
We are computing the Gegenbauer polynomials through recurrence, so the derivative is available to us for free through the formula
\begin{equation}
{C_n^{(\alpha)}}^\prime (x) = \frac{-n x C_n^{(\alpha)}(x)+(n+2 \alpha-1) C_{n-1}^{(\alpha)}(x)}{1-x^2},
\end{equation}
which can be derived from the recurrence relations satisfied by the Gegenbauer polynomials. Equation~\eqref{eq:CF1fromGegenbauer} will always work, but if $\nu \gg l$ it may be much faster to converge the continued fraction in equation~\eqref{eq:CF1}, depending on $
\chi$. If we suspect the continued fraction to be faster we try that first, but if it fails to converge we fall back on equation~\eqref{eq:CF1fromGegenbauer}. 
\subsection{Accuracy of the implementation}
\begin{figure}[tbp]
\begin{centering}
\includegraphics[width=\textwidth]{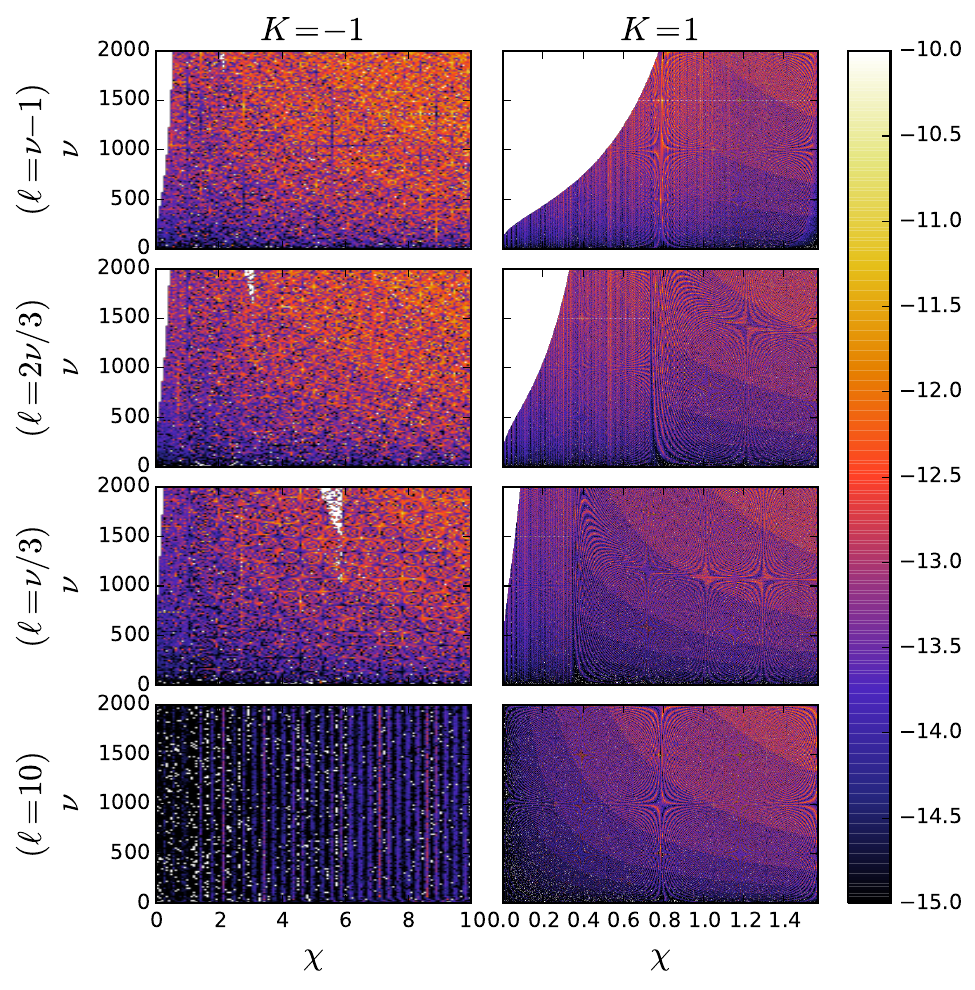}
\label{fig:reldif}
\caption{$\log_{10}$ of the magnitude of the relative difference between our implementation and an \mpmath{}-based implementation. In the connected white regions to the left of some subplots correspond to places where $\Phi$ vanishes to machine precision. The scattered white dots in the $K=-1$ plots are points where \mpmath{} failed to converge. The error is smaller than $10^{-12}$ for the while parameter space.}
\end{centering}
\end{figure}
We computed $\Phi_l^\nu(\chi)$ on a $(\chi,\nu)$-grid for $K=1$ and a $(\chi/\chi_\text{tp},\nu)$-grid for $K=-1$. Here $\chi_\text{tp}=\text{arcsinh}\left( \sqrt{l(l+1)}/\nu \right)$ denotes the value of the classical turning point which roughly corresponds to the location of the first peak. We fixed $l$ to 4 different cases: $\{l=10, l=\nu/3, l=2\nu/3, l= \nu=1\}$ and computed the relative error with respect to \mpmath{}. For $K=1$ we compared our result to a numerical implementation of equation~\eqref{eq:Gegenbauer}, and for $K=-1$ we relied on a combination of \mpmath{}'s \texttt{hyp2f1()} function and \texttt{legnp()} function. The result is displayed in figure~\ref{fig:reldif}, and as one can see the error is below $10^{-12}$ for the full parameter space.
\section{WKB approximation}
\begin{figure}[tbp]
\begin{centering}
\includegraphics[width=\textwidth]{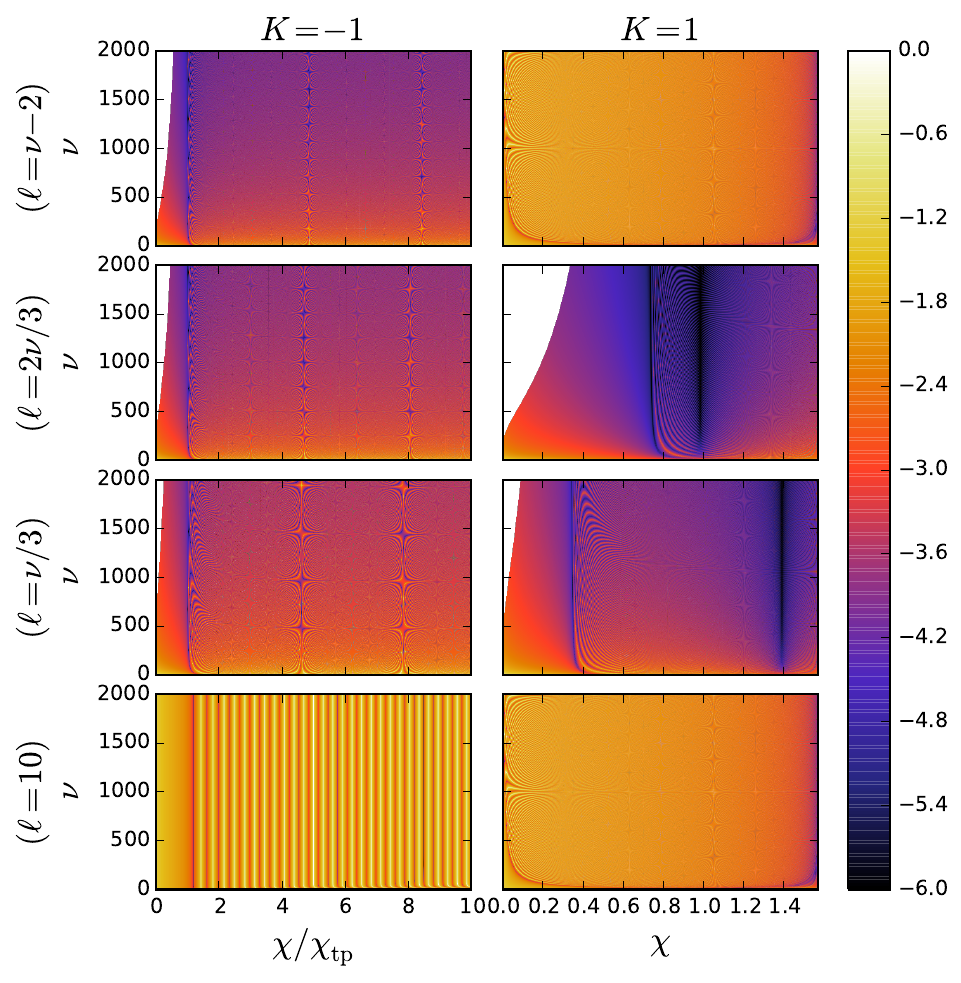}
\label{fig:reldifWKB}
\caption{$\log_{10}$ of the magnitude of the relative difference between the WKB approximation in equation~\eqref{eq:WKB} and our implementation. .}
\end{centering}
\end{figure}
Equation~\eqref{eq:u_equation} lends itself to a WKB approximation~\cite{Kosowsky:1998nc}. Using Langer's uniform approximation~\cite{Langer1935,bender1978advanced}, the WKB approximation reads
\begin{equation}
\Phi_l^\nu (\chi) \simeq \frac{\sqrt{\pi\alpha}}{\nu} Z^{\frac{1}{6}} \left|\frac{1}{\sinK^2\chi}-\alpha^2\right|^{-\frac{1}{4}}\frac{1}{\sinK \chi} \text{Ai}\(\text{sgn}(\chi_\text{tp}-\chi) Z^{\frac{2}{3}} \), \label{eq:WKB}
\end{equation}
 where $\text{Ai}(x)$ is the Airy function, $\alpha\equiv\frac{\nu}{\sqrt{l(l+1)}}$, and the turning point $\chi_\text{tp}$ is defined through $\sinK\chi_\text{tp}=\frac{1}{\alpha}$. $Z$ is given by
\begin{equation}
Z \equiv \frac{3}{2} S \sqrt{l(l+1)}, \qquad S \equiv \text{sgn}(\chi_\text{tp}-\chi) \int_{\chi_\text{tp}}^\chi \text{d}\chi' \sqrt{\left|\alpha^2-\frac{1}{\sinK^2\chi'} \right|},
\end{equation}
where the sign function is such that we always have $Z>S\geq0$. Defining $w\equiv \alpha \sinK\chi$, the definite integrals can be written in terms of elementary real functions\footnote{Similar formulae were given in~\cite{Kosowsky:1998nc}, but we disagree with 3 of the 4 formulae. The first one is identical to ours, while we believe there must be a typo in the second one. The third one has a constant offset for some range of $\chi$-values, likely due to a branch cut. Finally there is an $\alpha$ missing in front of the logarithm in the fourth formula.}:
\begin{align*}
\int_{\chi_\text{tp}}^\chi \text{d}\chi' \sqrt{\alpha^2-\frac{1}{\sinh^2\chi'} } &= \alpha \log\[\frac{\sqrt{w^2-1}+\sqrt{w^2+\alpha^2}}{\sqrt{1+\alpha^2}}\]+ 
   \atan\[\frac{1}{\alpha}\sqrt{\frac{w^2+\alpha^2}{w^2-1}}\]-\frac{\pi}{2}, \\
\int_\chi^{\chi_\text{tp}} \text{d}\chi' \sqrt{\frac{1}{\sinh^2\chi'}-\alpha^2}  &=  \atanh(u)-\alpha \atan\(\frac{u}{\alpha}\), \\
\int_{\chi_\text{tp}}^\chi \text{d}\chi' \sqrt{\alpha^2-\frac{1}{\sin^2\chi'} } &=  \atan(v)+\alpha \atan\(\frac{1}{v \alpha}\)-\frac{\pi}{2},\\
\int_\chi^{\chi_\text{tp}} \text{d}\chi' \sqrt{\frac{1}{\sin^2\chi'}-\alpha^2 } &= \atanh\[\frac{\sqrt{1-w^2}}{\sqrt{1-w^2/\alpha^2}}\]- 
\alpha \log\[\frac{\sqrt{\alpha^2-w^2}+\sqrt{1-w^2}}{\sqrt{\alpha^2-1}}\]. 
\end{align*}
Here we also defined $u\equiv \frac{\sqrt{1-w^2}}{\sqrt{1+\frac{w^2}{\alpha^2}}}$ and $v=\frac{\sqrt{1-\frac{w^2}{\alpha^2}}}{\sqrt{w^2-1}}$. For $K=-1$, the formulae are valid for $\chi>0$, while for $K=1$ they are valid for $0<\chi<\frac{\pi}{2}$. However, as discussed earlier this is sufficient because we can extend the solution to the whole real axis through equation~\eqref{eq:evenodd} and equation~\eqref{eq:symmetry}. When deriving these formulae from the indefinite integral, one must be careful in avoiding branch cuts of the complex functions.

We have computed the relative error of the WKB approximation to our recurrence-based implementation in figure~\ref{fig:reldifWKB}. We have taken the same parameter space as in figure~\ref{fig:reldif} except that we have substituted $l=\nu-1$ for $l=\nu-2$. This is because the WKB approximation for $K=1$ is known to be inaccurate for that particular value~\cite{Kosowsky:1998nc} so it is not representative. $l=\nu-2$ suffers to some extent from the same problem as can be seen from the plot. For $l=10$ the error is the \%-level, while for larger $l$ the error drops below $10^{-3}$.

In contrast to what one may have expected, our implementation of the WKB approximation turned out to be significantly slower than the recurrence method when multiple $l$-values were required. This is because each point will require two trigonometric function calls, at least one squareroot and the value of the Airy function. 
\section{Conclusion}\label{sec:conclusion}
We have developed fast and accurate algorithms for computing hyperspherical Bessel functions for real arguments and possibly large orders. While similar methods for $K=-1$ has been available for some time, the $K=1$ case was never implemented satisfactorily due to the problem of backwards recurrence. We solved this problem by exploiting an identity between Legendre functions and Gegenbauer functions to derive an identity between the $K=1$ hyperspherical Bessel functions and the Gegenbauer polynomials.

All routines are available as part of the public CMB-code~\CLASS{}\footnote{Webpage at \url{http://class-code.net} and public GitHub repository at \url{http://github.com/lesgourg/class_public}.} written in C. All routines related to the hyperspherical Bessel functions are in a separate file, \texttt{hyperspherical.c} so it can easily be extracted from~\CLASS{}. In fact, the only dependency is a set of macros defined in the file \texttt{common.h}. A Python wrapper for the hyperspherical Bessel functions is available upon request.

\bibliographystyle{JHEP}

\bibliography{hyperbessel}

\end{document}